\documentclass[12pt]{article}
\newcommand{\beq}{\begin{equation}}
\newcommand{\eeq}{\end{equation}}
\newcommand{\bra}{\begin{array}}
\newcommand{\era}{\end{array}}
\newcommand{\be}{\beta}
\newcommand{\te}{\theta}
\newcommand{\al}{\alpha}

\newcommand{\de}{\delta}
\newcommand{\da}{\dagger}

\newcommand{\om}{\omega}

\newcommand{\ep}{\epsilon}
\usepackage{latexsym}
\author{Jamila Douari \footnote{douari@sun.ac.za} \\
\small\it Stellenbosch Institute for Advanced Study, Private Bag X1,\rm\\
\small\it Matieland, Stellenbosch, 7601, South Africa\rm }
\title{Deformed $C_{\lambda}$-Extended Heisenberg Algebra in Non-commutative Phase-Space}
\begin{document}
\maketitle
\vspace*{0.5cm}
\hspace*{1.1in}.
\vspace*{2cm}
\section*{Abstract}
\hspace{.3in}We construct a deformed $C_{\lambda}$-extended Heisenberg algebra in two-dimensional space using non-commuting coordinates which close an algebra depends on statistical parameter characterizing exotic particles. The obtained symmetry is nothing but an exotic particles algebra interpolating between bosonic and deformed fermionic algebras.
\section{Introduction}
\hspace{.3in}The recent upsurge of interest in the physics in non-commutative spaces has been spurred due to its very clear and cogent appearance in the context of string theories, D-branes and M-theories. In quantum field theory, this is motivated by studies of the low energy effective theory of D-brane with a non-zero NS-NS B field background \cite{ncst}. On other hand, the study at the level of quantum mechanics in non-commutative spaces is also meaningful for clarifying some possible phenomenological consequences in solvable models. In this latter context, the present paper is basically devoted to deal with exotic particles living in two-dimensional space, such that a consistent ansatz of commutation relations of phase-space variables should simultaneously include space-space non-commutativity and momentum-momentum non-commutativity. Then, we obtain a new type of commutation relations at the deformed level defining the exotic particles algebra and we show that the obtained symmetry is nothing but a deformed $C_{\lambda}$-extended Heisenberg algebra which is kind of deformed oscillator algebra. As known in the literature, the subject of oscillator algebras has been a topic of intensive research activities during the past decade and much attention has been paid to this field in connection with the important role investigated in many physical systems, such as the description of the fractional statistics \cite{b,any,symany}. Among these various deformations and extensions, we mention the generalized deformed oscillator algebras (GDOA's) \cite{c} and the $G$-extended oscillator algebras \cite{d} where $G$ is some finite group, e.g in the case of Calogero-model $C_{\lambda}=\mathcal{Z}_{\lambda}$ is the cyclic group of order $\lambda$.

The goal of the present letter is mainly to obtain an algebra describing the planar system (exotic particles), which is different from that given by Lerda and Sciuto in the ref. \cite{as}. The study is based on the non-commutative geometry defined by a fundamental algebra depending on the statistical parameter $\nu$ which characterizes exotic particles. In section 2, we give a short review on these latter particles. Then we recall some facts concerning the $C_{\lambda}$-extended oscillator algebras which were introduced as a generalization of Calogero-Vasiliev algebras \cite{c,d,e} in section 3. In section 4, we consider a non-commuting spatial and momentum coordinates satisfying an algebra depending on the statistical parameters to define an annihilation and creation operators which generate an exotic particles algebra. Owing to the latter algebra. We also show that the obtained symmetry is interpolating between bosonic and deformed fermionic one for arbitrary operator $\xi$ which is introduced to define the generators. Another important result is that the algebra describing the planar system is a defomed $C_\lambda$-extended Heisenberg algebra. 

\section{Anyons}
\hspace{.3in}These particles are known as quasi-particles or excitations \cite{lauph} in two-dimensional space obey intermediate statistics that interpolate between bosonic and fermionic statistics because of its multiply connected configuration space \cite{b}. Among the theories describing anyons there is Chern-Simons theory \cite{any} and generalized Maxwell theory \cite{GConn}. In the first theory, the standard way to obtain anyons is to add to the action $S$ a topological, or Hopf, term
\beq
S\longrightarrow S+i\nu S_{top}.
\eeq
The simplest case, to describe free anyons, corresponds to $S$ being free charged bosons or fermions with U(1) current $J^\mu$ and charges $Q = \pm 1$. Here $S_{top}$ is the Chern-Simons term where the $U(1)$ gauge field $A_\mu$ is manufactured from the U(1) matter current $J^\mu=\ep^{\mu\nu\al}\partial_\nu A_\al$. The statistical parameter $\nu$ is normalized so that the spin of the
particle is $\frac{\nu}{2\pi}$. The second theory is a novel way to describe anyons without a Chern-Simons term. Thus, a generalized connection was considered in (2+1)-dimensions denoted $A_\mu^\te$, $\mu=0,1,2$. The gauge theory is defined by the following Lagrangian
\beq
L_\te=-\frac{1}{2}F_{\mu\nu}F^{\mu\nu}+J^\mu A_\mu^\te
\eeq
with $A_\mu^\te\equiv A_\mu +\frac{\te}{2}\ep_{\mu\nu\rho}F^{\nu\rho}$ is the generalized connection and $\te$ is real parameter in Minkowski space. The Lagrangian $L_\te$ desribes Maxwell theory that couples to the current via the generalized connection rather than the usual connection. In this model, the gauge field is dynamic and the potential has confining nature which make the theory different from the first one.

On other hand, at the level of quantum mechanics in non-commutative spaces, the noncommutative geometry and anyons are related as was shown in the references \cite{annc}. The noncommutativity comes from the presence of the magnetic field. In this sense it is valuable to study a field theory both in the presence of a magnetic field and in the coordinate space. In this context, the present work is devoted to find out the symmetry describing anyons basing on noncommutative geometry. We found it very interesting by considering the algebra closed by the coordinates depends on statistical parameter as we will see in the next section. The obtained algebra is interpolating between bosonic and deformed fermionic algebras depending on statistical parameter. 

After this short review on planar system, we give in what follows its associated symmetry having two extremes: bosonic and deformed fermionic symmetries.

\section{Planar System Symmetry}
\hspace*{.3in}First, we briefly recall the non-commutative geometry. Its most simple example consists of the geometric space described by non-commutative hermitian operator coordinates $x_{i}$, and by considering the non-commutative momentum operators $p_i=i\partial_{x_i}$ $(\partial_{x_i}$ the corresponding derivative of $x_i$). These operators satisfy the following algebra
\beq
\bra{cc}
\lbrack x_{i} ,x_{j} \rbrack = i\theta\epsilon_{ij},\phantom{~~~~~}
\lbrack p_i,p_j \rbrack =i\theta^{-1}\epsilon_{ij},\phantom{~~~~~}\lbrack p_i,x_j \rbrack =-i\delta_{ij}\\\\
\lbrack p_i ,t\rbrack =0=\lbrack x_i ,t\rbrack ,\phantom{~~~~~}\lbrack p_i,\partial_{t}\rbrack =0=\lbrack x_i,\partial_{t}\rbrack,
\era
\eeq
with $t$ the physical time and $\partial_{t}$ its corresponding derivative.

By considering two-dimensional harmonic oscillator which can be decomposed into one-dimensional oscillators. So, it is known that the algebra (2) allows to define, for each dimension, the representation of annihilation and  creation operators as follows
\beq
\bra{ll}
a_{i}&=\sqrt{\frac{\mu\om}{2}}(x_i +\frac{i}{\mu\om} p_i ) \\
a_{i}^{\da}&=\sqrt{\frac{\mu\om}{2}}(x_i -\frac{i}{\mu\om}  p_i ) .\\
\era
\eeq
with $\mu$ is the mass and $\om$ the frequency. These operators satisfy $$\lbrack a_i ,a^{\da}_i \rbrack =1,$$ defining the Heisenberg algebra. In the simultanuously non-commutative space-space and non-commutative momentum-momentum, the bosonic statistics should be maintained; i.e, the operators $a_{i}^{\da}$ and $a_{j}^{\da}$ are commuting for $i\ne j$. Thus, the deformation parameter $\te$ is required to satisfy the condition $$\te=-(\frac{1}{\mu\om})^2\te^{-1}.$$

To find out an algebra describing the planar system we start by introducing the non-commutative geometry depending on the statistical parameter $\nu\in\bf R\rm$. We study two cases with the main difference is based on the deformation of the commutative relation \beq \lbrack p_i ,x_j \rbrack =-i\delta_{ij}.\eeq 
\subsection{Case 1}
In this case, we deform the commutation (5). Thus, the fundamental algebra is defined by the spatial coordinates $x_i$ and the momentum $p_i$ satisfying\\

{\bf Proposition 1}
\beq
\bra{cc}
\lbrack x_i ,x_j \rbrack_\chi = i\theta \epsilon_{ij},\phantom{~~~~~} \lbrack p_i ,p_j \rbrack_\chi =- i\theta (\mu\om)^{2}\epsilon_{ij},\phantom{~~~~~} \lbrack p_i ,x_j \rbrack_\chi =-i\eta\delta_{ij}\\\\
\lbrack p_i ,t\rbrack =0=\lbrack x_i ,t\rbrack ,\phantom{~~~~~}\lbrack p_i,\partial_{t}\rbrack =0=\lbrack x_i,\partial_{t}\rbrack,
\era
\eeq
with the second deformation parameter $\chi$ is given by\\

{\bf Definition 1}
\beq
\chi=e^{\pm i\nu\pi},
\eeq
where $\pm$ sign indicates the two rotation directions on two-dimensional space. $\te$ and $\eta$ are a non-commutative parameters depending on statistical parameter $\nu$ as we will see later and the notation $[x,y]_q =xy-qyx$. Thus, owing to the third equality in (6) we get $$\lbrack x_i ,p_j \rbrack_\chi =(\chi^{-1}-\chi)p_j x_i+i\chi^{-1}\eta\delta_{ij}.$$

Then, we introduce an operator $\xi_i$ acting on the momentum direction in the phase-space. We assume that $\xi_i$ satisfies the following commutation relation\\

{\bf Proposition 2}
\beq
\lbrack\xi_i ,x_{j}\rbrack =0\phantom{~~~~~}\forall i,j.
\eeq
In this case, we define the annihilation and the creation operators by\\

{\bf Definition 2}
\beq
\bra{ll}
b_{i}^-&= \sqrt{\frac{\mu\om}{2}}(x_i +\frac{i}{\mu\om} \xi_i p_i ) \\ \\
b^+_{i}&= \sqrt{\frac{\mu\om}{2}}(x_i -\frac{i}{\mu\om} \xi^{-1}_i p_i ) ,
\era
\eeq
with $\xi_i $ is defined in terms of statistical parameter $\nu$ and an operaor $K_i $ which could be a function of the number operator $N$\\

{\bf Definition 3}
\beq
\xi_i =e^{i\nu\pi K_i },
\eeq

According to (6-7) the non-commutative geometry leads to a deformed Heisenberg algebra satisfied by the operators (9) and defined by the following commutation relations
\beq
\bra{lll}
\lbrack b^-_{i},b^+_{j} \rbrack_\chi &= \frac{1}{2}\eta(\xi_i \chi^{-1}+\xi^{-1}_j ) \de_{ij}+i\frac{\mu\om}{2}\te(I+\xi_i \xi^{-1}_j)\epsilon_{ij}-\frac{i\xi_j^{-1}}{2}B_{ij},\\ \\
\lbrack b^+_{i},b^+_{j} \rbrack_\chi &= \frac{1}{2}\eta(\xi^{-1}_j -\xi^{-1}_i \chi^{-1}) \de_{ij}+ i\frac{\mu\om}{2}\te(I-\xi^{-1}_i \xi^{-1}_j ) \epsilon_{ij}-\frac{i\xi_j^{-1}}{2}B_{ij},\\ \\
\lbrack b^-_{i},b^-_{j} \rbrack_\chi &= \frac{1}{2}\eta(\xi_i \chi^{-1}-\xi_j ) \de_{ij}+i\frac{\mu\om}{2}\te(I-\xi_i \xi_j )\epsilon_{ij}+\frac{i\xi_j}{2}B_{ij},
\era
\eeq
with $I$ is the identity and $B_{ij}=(\chi^{-1}-\chi)p_j x_i$.

To be consistent with the hermiticty of coordinates we suggest that $\te$ and $\eta$ are operators satisfying $$\theta^\dagger=\chi^{-1}\theta,\phantom{~~~~~}\eta^\dagger=\chi^{-1}\eta$$ and we assume that there is some function $f:\bf R\rm \longrightarrow \bf R\rm$ such that $$\bra{rcl} f^\dagger (\nu)=f(\nu),& f(0)\longrightarrow1 \era$$ and $$\lim\limits_{\nu\longrightarrow 1} \frac{1+\chi}{f(\nu)} =2.$$ Then, we give the following expressions\\

{\bf Definition 4}
\beq
\bra{ll}
\te =\nu(1+\chi)I\\\\
\eta=\frac{1}{2}\frac{1+\chi}{f(\nu)}\cos\nu 2\pi,
\era
\eeq
which are compatible with the commutative phase-space at the extremes concerning the statistical parameter $\nu$; $\te=0$ and $\eta= 1$ for $\nu=0,1$ respectively. We remark that the algebra (11) is a deformed version of Heisenberg algerbra satisfied by the operators given in (3). This new algebra describes the anyonic system for arbitrary statistical parameter $\nu$.

Another important point is that the obtained deformed Heisenberg algerbra (11) is interpolating between two extremes depending on  the statistical parameter $\nu$. We know that, in three or more dimensions, $\nu$ takes the values 0 or 1 and in two dimensions $\nu$ is arbitrary real number. The latter case has already discussed above and characterizing exotic particles. In the case of three or more dimensions if $\nu=0$ we get $\chi=1$, $\xi_i =\xi^{-1}_i =I$, $\te=0$, $\eta=1$ and $B_{ij}=0$. Thus, the commutation relations of the algebra (11) becomes
\beq
\bra{lcr}
\lbrack b^-_{i},b^+_{j} \rbrack = \de_{ij},&
\lbrack b^+_{i},b^+_{j} \rbrack = 0,&
\lbrack b^-_{i},b^-_{j} \rbrack = 0.
\era
\eeq
These relations define the bosonic algebra and this is one extreme. The second extreme could be gotten if $\nu=1$, then we have $\chi=-1$, $\te=0$,  $\eta=1$ and  $B_{ij}=0$. We find the following commutation relations
\beq
\bra{lll}
\lbrace b^-_{i},b^+_{j} \rbrace &= \frac{1}{2}( e^{-i\pi K_j}-e^{i\pi K_i})\de_{ij},\\\\
\lbrace b^+_{i},b^+_{j} \rbrace &= \frac{1}{2}(e^{-i\pi K_j} +e^{-i\pi K_i})\de_{ij},\\\\
\lbrace b^-_{i},b^-_{j} \rbrace &= \frac{1}{2}(e^{i\pi K_i} +e^{i\pi K_j})\de_{ij}.
\era
\eeq
which close a deformed fermionic algebra for arbitrary operator $K_i$ (10) as a second extreme for the algebra (11). Again if $K_i $ is a hermitian operaor then $\xi_i$ is unitary and $b^+_{i}$ is a complex conjugate of $b_{i}^-$, then the algebra (11) will not have the fermionic algebra as extreme when $\nu=1$ but its two extremes are bosonic algebra and deformed fermionic algebra.
\subsection{Case 2}
\hspace{.3in}In this subsection, let us see what will happen if we don't deform the combined commutator of momentum and spatial coordinates. If we keep the commutation relation (5) plus the other relations of the equation (6) the noncommutative geometry is now defined by the following fundamental algebra\\

{\bf Prposition 3}
\beq
\bra{cc}
\lbrack x_i ,x_j \rbrack_\chi = i\theta \epsilon_{ij},\phantom{~~~~~} \lbrack p_i ,p_j \rbrack_\chi =- i\theta (\mu\om)^{2}\epsilon_{ij},\phantom{~~~~~}\lbrack p_i ,x_j \rbrack_ =-i\delta_{ij} \\\\
\lbrack p_i ,t\rbrack =0=\lbrack x_i ,t\rbrack ,\phantom{~~~~~}\lbrack p_i,\partial_{t}\rbrack =0=\lbrack x_i,\partial_{t}\rbrack.
\era
\eeq
By straightforward calculations we obtain
\beq
\bra{lr}
\lbrack x_i ,p_j\rbrack_\chi =i\de_{ij}+C_{ji},&
\lbrack p_i ,x_j \rbrack_\chi =-i\de_{ij}+D_{ji},
\era
\eeq
where $C_{ji}=(1-\chi)p_j x_i$ and $D_{ji}=(1-\chi)x_j p_i$.\\

Consequently the exotic particles algebra becomes
\beq
\bra{lll}
\lbrack b^-_{i},b^+_{j} \rbrack_\chi &= \frac{1}{2}(\xi_i +\xi^{-1}_j ) \de_{ij}+i\frac{\mu\om}{2}\te(I+\xi_i \xi^{-1}_j)\epsilon_{ij} -\frac{i}{2}(\xi_j^{-1}C_{ji}-\xi_i D_{ji}),\\ \\
\lbrack b^+_{i},b^+_{j} \rbrack_\chi &= \frac{1}{2}(\xi^{-1}_j -\xi^{-1}_i ) \de_{ij}+ i\frac{\mu\om}{2}\te(I-\xi^{-1}_i \xi^{-1}_j ) \epsilon_{ij} -\frac{i}{2}(\xi_j^{-1}C_{ji}+\xi_i^{-1}D_{ji}),\\ \\
\lbrack b^-_{i},b^-_{j} \rbrack_\chi &= \frac{1}{2}(\xi_i -\xi_j ) \de_{ij}+i\frac{\mu\om}{2}\te(I-\xi_i \xi_j )\epsilon_{ij}+\frac{i}{2}(\xi_j C_{ji}+\xi_i D_{ji}),
\era
\eeq
with $\te$ is defined by (12). Again, if $\nu=0$ we get $\chi=1$, $\te=0$ and $C_{ji}=0=D_{ji}$ and we refind the bosonic algebra (13) as an extreme of the symmetry (17) describing quasi-particles system. Then the case $\nu=1$ leads to $\chi=-1$, $\te=0$ and $C_{ji}\ne 0\ne D_{ji}$ and we find a deformed fermionic algebra which is different from (14) and defined by
\beq
\bra{ccc}
\lbrace b^-_{i},b^+_{j} \rbrace = \frac{1}{2}(e^{i\pi K_i} +e^{-i\pi K_j} ) \de_{ij} -\frac{i}{2}(e^{-i\pi K_j}C_{ji}-e^{i\pi K_i} D_{ji}),\\ \\
\lbrace b^+_{i},b^+_{j} \rbrace = \frac{1}{2}(e^{-i\pi K_j} -e^{-i\pi K_i} ) \de_{ij}-\frac{i}{2}(e^{-i\pi K_j}C_{ji}+e^{-i\pi K_i}D_{ji}),\\ \\
\lbrace b^-_{i},b^-_{j} \rbrace = \frac{1}{2}(e^{i\pi K_i} -e^{i\pi K_j} ) \de_{ij}+\frac{i}{2}(e^{i\pi K_j} C_{ji}+e^{i\pi K_i} D_{ji}).
\era
\eeq
as a second extreme of (17).

The main result we get from this investigation is that exotic particles algebra goes to bosonic algebra if $\nu\longrightarrow 0$. This means that our system is originally gotten by exciting a bosonic system in two-dimensional space. Also, we get a deformed fermionic algebra as a second extreme when the statistical parameter $\nu$ equals to 1. Thus, we remark that the system described by the above algebras (11) or (17) doesn't have any thing to do with fermions originally but it could be related to something else as deformed fermions which are known in the literature as quionic particles or $k_i$-fermions, $k_i$ integer number introduced as deformation parameter, and these kinds of particles are known as non physical particles.
\section{Deformed Oscillator Algebras}
\hspace*{.3in}In this section we show that the extended Heisenberg algebra could be a symmetry of planar system at defomed level. First we start by a short review on $C_{\lambda}$-extended oscillator algebra and then we give the defomed form of this symmetry which describes exotic particles in two-dimensional space.
\subsection{Extended Heisenberg Algebra}
\hspace*{.3in}We review in brief the $C_{\lambda}$-extended oscillator algebras. As known in the literature, a generalization of the Calogero-Vasiliev algebras, the $C_{\lambda}$-extended oscillator algebras (also
called generalized deformed oscillator algebras (GDOA's)), denoted $A^{\lambda}$, $\lambda=2,3,...,$ are defined by
\beq \bra{lrlr}
[N,a^{\dagger}]=a^{\dagger},& [a,a^{\dagger}]=I+\sum\limits_{\mu=0}^{\lambda-1}\alpha_{\mu}P_{\mu},\\ \\

[N,P_{\mu}]=0,& a^{\dagger}P_{\mu}=P_{\mu +1}a^{\dagger},
\era \eeq together with their hermitian conjugates, and \beq
\bra{lrlr}
P_{\mu}=\frac{1}{\lambda}\sum\limits_{\nu=0}^{\lambda-1}
e^{\frac{2\pi i\nu(N-\mu)}{\lambda}},&
\sum\limits_{\mu=0}^{\lambda-1}P_{\mu}=1,&
P_{\mu}P_{\nu}=\de_{\mu,\nu}P_{\nu}\\ \\
\sum\limits_{\mu=0}^{\lambda-1}\alpha_{\mu}=0,&
\sum\limits_{\nu=0}^{\mu-1}\alpha_{\nu}>-1,& \mu=1,...,\lambda-1.
\era, \eeq  where $\alpha_{\mu}\in{\bf R\rm}$, $N$ is the number
operator and $P_{\mu}$ are the projection operators on subspaces
$F_{\mu}=\{\vert k\lambda -\mu\rangle \vert k=0,1,2,...\}$ of the
Fock space $F$ which is portioning into $\lambda$ subspaces.

The operators $a$ and $a^{\dagger}$ are defined by \beq \bra{lrlr}
a^{\dagger}a=F(N),& aa^{\dagger}=F(N+1), \era \eeq where
$F(N)=N+\sum\limits_{\mu=0}^{\lambda-1}\be_{\mu}P_{\mu}$,
$\be_{\mu}=\sum\limits_{\nu=0}^{\mu-1}\al_{\nu}$, which is a
fundamental concept of deformed oscillators. Let's  denote the
basis states of subspaces $F_{\mu}$ by $\vert n\rangle=\vert
k\lambda +\mu\rangle\simeq (a^{\dagger})^n\vert 0\rangle$ where
$a\vert 0\rangle=0$, $\vert 0\rangle$ is the vacuum state. The
operators $a$, $a^{\dagger}$  and $N$ act on $F_{\mu}$ as
follows \beq \bra{rcl} N\vert n\rangle=n\vert n\rangle,&
a^{\dagger}\vert n\rangle=\sqrt{F(N+1)}\vert n+1\rangle,& a\vert
n\rangle=\sqrt{F(N)}\vert n-1\rangle. \era \eeq According to these
relations, $a$ and $a^{\dagger}$ are the
annihilation and the creation operators respectively.

Particularly, if $\lambda=2$, we have two projection operators
$P_{0}=\frac{1}{2}(I+(-1)^{N})$ and
$P_{1}=\frac{1}{2}(I-(-1)^{N})$ on the even and odd subspaces of
the Fock space $F$, and the relations of (1) are restricted to
\beq \bra{rcl} [N,a^{\dagger}]=a^{\dagger},&
[a,a^{\dagger}]=I+\kappa K,& \{K,a^{\dagger}\}=0, \era \eeq with
their hermitian conjugates, where $K=(-1)^{N}$ is the Klein
operator and $\kappa$ is a real parameter. These relations define
the so-called Calogero-Vasiliev algebra.

The $C_{\lambda}$-extended oscillator algebras are seeing as
deformation of $G$-extended oscillator algebras, where $G$ is some
finite group, appeared in connection with $n$-particle integrable
models. In the former case, $G$ is the symmetric group $S_n$. So,
for two particles $S_2$, can be realized in terms of $K$ and
$S_2$-extended oscillator algebra becomes a generalized
deformed oscillator algebra (GDOA) also known as the
Calogero-Vasiliev or modified oscillator algebra. In the
$C_{\lambda}$-extended oscillator algebras, $G\equiv C_{\lambda}$
is the cyclic group of order $\lambda$,
$C_{\lambda}=\{1,K,...,K^{\lambda-1}\}$. So, these algebras have a
rich structure since they depend upon $\lambda$ independent real
parameters, $\alpha_0 ,\alpha_1
,...,\alpha_{\lambda-1},$.
\subsection{Deformed $C_{\lambda}$-Extended Heisenberg Algebra}
\hspace*{.3in}Other version of anyonic algebra can be obtained by treating the special case of statistical parameter $\nu\in[0,1]$. By using, the Taylor expansion to the operator $\xi_i $, the first commutation relation in the algebra (17) can be rewritten in this form
\beq
\lbrack b^-_{i},b^+_{j} \rbrack_\chi =(I+\Re_{ij}^{[\nu]})\de_{ij}+A^{[\nu]}_{ij}\ep_{ij}+Q^{[\nu]}_{ij},
\eeq
where
\beq
\Re_{ij}^{[\nu]}=\sum\limits_{\ell=1}^{\frac{n+1}{2}}\kappa_{\nu,\ell}\frac{K_i ^{2\ell -1}-K_j ^{2\ell -1}}{2}+\sum\limits_{k=1}^{\frac{m}{2}}\kappa_{\nu,k}\frac{K_i ^{2k }+K_j ^{2k}}{2},
\eeq
$$Q^{[\nu]}_{ij}=-\frac{i}{2}\sum\limits_{p=0}^{\lambda -1}\frac{(i\nu\pi)^p }{p!}((-K_j)^{p}C_{ji}-K_i^p D_{ji})$$
and 
\beq
A^{[\nu]}_{ij}=\frac{i\te\mu w}{2}\Big( I+\sum\limits_{\al=0}^{\lambda -1}\frac{(i\nu\pi)^\al }{\al!}(K_i -K_j )^\al\Big) ,
\eeq
with $m\in\bf N \rm$ is even and $n\in\bf N \rm$ odd such that $n,m\le\lambda -1$ with $\lambda\in\bf N \rm$ by imposing $$K_i ^{\lambda} = I.$$ 
 The coeffecients $\kappa_{\nu,\ell}$ and $\kappa_{\nu,k}$ are given in terms of statistical parameter as follows
$$\kappa_{\nu,\ell}=\frac{(i\nu\pi)^{2\ell -1}}{(2\ell -1)!},\phantom{~~~~}\kappa_{\nu,k}=\frac{(i\nu\pi)^{2k}}{(2k)!}.$$
Then, the last two commutation relations of (17) become
\beq
\bra{ll}
\lbrack b^+_{i},b^+_{j} \rbrack_\chi &=  \sum\limits_{\al=1}^{\lambda -1}\frac{(-i\nu\pi)^\al }{\al!}\frac{K_j^\al -K_i^\al }{2} \de_{ij}-i\frac{\mu\om}{2}\te\sum\limits_{\al=1}^{\lambda -1}\frac{(-i\nu\pi)^\al }{\al!}(K_i +K_j )^\al \epsilon_{ij}\\\\&-
\frac{i}{2}\sum\limits_{p=0}^{\lambda -1}\frac{(-i\nu\pi)^p }{p!}(K_j^{p}C_{ji}-K_i^p D_{ji}),\\ \\
\lbrack b^-_{i},b^-_{j} \rbrack_\chi &=  \sum\limits_{\al=1}^{\lambda -1}\frac{(i\nu\pi)^\al }{\al!}\frac{K_i^\al -K_j^\al }{2} \de_{ij}-i\frac{\mu\om}{2}\te\sum\limits_{\al=1}^{\lambda -1}\frac{(i\nu\pi)^\al }{\al!}(K_i +K_j )^\al \epsilon_{ij}\\\\&+\frac{i}{2}\sum\limits_{p=0}^{\lambda -1}\frac{(i\nu\pi)^p }{p!}(K_j^{p}C_{ji}-K_i^p D_{ji})
.
\era
\eeq
Now if we pose the following\\

{\bf Proposition 4}
$$K_i =e^{i\frac{2\pi}{\lambda}N_i }$$
with $N_i $ is a number operator defined in terms of $b^+_{i}$ and $b^-_{i}$ as
\beq
b^+_{i}b^-_{i}=f(N),\eeq
with $f$ is some function such that the following relations are satisfied
$$\bra{ll}
\lbrack N_i, b^-_{j}\rbrack = -\de_{ij}b_{i},\phantom{~~~~~~}\lbrack N_i, b^+_{j}\rbrack = \de_{ij}b^{\da}_{i},\\\\
K_i b_{j} = \de_{ij}e^{-i\frac{2\pi}{\lambda}}b_{j}K_i ,\phantom{~~~~~~}K_i b^{\da}_{j} = \de_{ij}e^{i\frac{2\pi}{\lambda}}b^{\da}_{j}K_i.
\era
$$

Thus, It is easy to see that the obtained relations (24) and (27) define a physicswise realization of deformed $C_{\lambda}$-extended Heisenberg algebra on two-dimensional non-commutative space describing exotic particles, where $C_{\lambda}$ is a cyclic group $$C_{\lambda}=\{I, K_i , K^2_i ,K^3_i ,...,K^{\lambda-1}_i \},\phantom{~~~~}\lambda\in\bf N\rm.$$ Again once, it is clear that the obtained algebra goes to bosonic symmetry if $\nu$ goes to 0.
\subsection{Fock Representation}
\hspace{.3in}It is convenient to construct a Fock representation for the algebra (15) underlying the non-commutative geometry by way of the operators (9) obeying (24,27). The Fock space is introduced by the set
\beq
F_i=\{\mid n\rangle;\phantom{~~}n=0,1\}
\eeq
with the states $\mid n\rangle$ are defined as
\beq
\mid n\rangle=\frac{1}{f(\sqrt{n!})}(b_i^+ )^n \mid 0\rangle,\phantom{~~~~}n=0,1
\eeq
they are the quantum mechanical states inherent to the non-commutativity (15), with $\mid 0\rangle$ the vacuum state and $f$ is some funtion definning the nubmer operator $N_i$ as given above in (28). The "exotic" annihilation and creation operators act on Fock space as
\beq\bra{ll}
b^+_{i}\mid n\rangle =f(\sqrt{n+1}) \mid n+1\rangle,\\\\
b^-_{i}\mid n\rangle =f(\sqrt{n}) \mid n-1\rangle.
\era\eeq
Then, Owing to (9,31), the non-commuting spatial and momentum coordinates are acting on Fock space as
\beq\bra{ll}
x_i \mid n\rangle =\frac{1}{\sqrt{2\mu w}}\Big[ f(\sqrt{n}) \mid n-1\rangle+ f(\sqrt{n+1}) \mid n+1\rangle \Big],\\\\
p_i \mid n\rangle =\frac{i\sqrt{2\mu w}}{e^{i\frac{2\pi}{\lambda}n }-e^{-i\frac{2\pi}{\lambda}n }}\Big[ -f(\sqrt{n}) \mid n-1\rangle+ f(\sqrt{n+1}) \mid n+1\rangle \Big],
\era\eeq
\section{Conclusion}
\hspace{.3in}We want to mention in conclusion that one of the important results we got studing the noncommutative geometry of two-dimensional space is the crucial role of statistical parameter $\nu$. In this investigation, we deformed the fundamental algebra describing the noncommutative geometry to find out the symmetry describes exotic particles living in two-dimensional space. The study leaded to an algebra interpolating between bosonic and deformed fermionic algebras. This means that our system is originally gotten by exciting a bosonic system in two-dimensional space and since the second extreme is a deformed fermionic algebra, the exotic particles system doesn't have anything to do with fermions originally when the statistical parameter $\nu$ goes to 1 but it could be related to something else as deformed fermions which are known in the literature as quionic particles or $k_i$-fermions, $k_i$ integer number introduced as deformation parameter, and these kinds of particles are not physical particles. Then we looked for other face of exotic particles algebra characterized by small statistical parameters. We obtained a deformed $C_{\lambda}$-Extended Heisenberg Algebra describing quasi-particles. This result is a realization, in physicswise, of a deformed version of $C_{\lambda}$-Extended Heisenberg Algebra \cite{eha} in two-dimensional non-commutative space.\\

Acknowledgements: The author would like to thank the Abdus Salam ICTP for the hospitality during the visit in which a part of this work was done.

\end{document}